\documentclass[aps,pre,twocolumn,nofootnoteinbib,superscriptaddress]{revtex4-2}
\usepackage{epsf,amsmath,amssymb,verbatim,color,multirow,pifont,graphicx,cleveref,tabularx,cancel,soul,ulem}

\newcommand{\Nex}{N_{\rm ex}}
\newcommand{\Nim}{N_{\rm im}}
\newcommand{\Npr}{N_{\rm pr}}

\newcommand{\kcnn}{k_{{\rm nn},c}}
\newcommand{\rcnn}{r_{{\rm nn},c}}

\newcommand{\qnn}{q_{\rm nn}}
\newcommand{\knn}{k_{\rm nn}}
\newcommand{\Snn}{S_{\rm nn}}
\newcommand{\rnn}{r_{\rm nn}}
\newcommand{\Unn}{U_{\rm nn}}

\newcommand{\qEI}{q^{\rm (E-I)}}
\newcommand{\kEI}{k^{\rm (E-I)}}

\newcommand{\qEP}{q^{\rm (E-P)}}
\newcommand{\rEP}{r^{\rm (E-P)}}

\newcommand{\thetaex}{\theta^{\rm (ex)}}
\newcommand{\thetaim}{\theta^{\rm (im)}}
\newcommand{\thetapr}{\theta^{\rm (pr)}}

\newcommand{\lambdaex}{\lambda^{\rm (ex)}}
\newcommand{\lambdaim}{\lambda^{\rm (im)}}
\newcommand{\lambdapr}{\lambda^{\rm (pr)}}

\newcommand{\aERH}{a^{\rm ERH}}
\newcommand{\aEI}{a^{\rm (E-I)}}
\newcommand{\aEP}{a^{\rm (E-P)}}
\newcommand{\wEI}{w^{\rm (E-I)}}
\newcommand{\wEP}{w^{\rm (E-P)}}
\newcommand{\aCH}{a^{\rm CH}}
\newcommand{\acorr}{a^{ \rm COR}}

\newcommand{\Normc}{\mathcal{N}_c}
\newcommand{\wEPpnn}{w^{\rm (E-P)}_{{\rm nn},p}}

\newcommand{\Enn}{E_{\rm nn}}

\begin{document}
\title{Structure of international trade hypergraphs}
\author{Sudo  \surname{Yi}}
\affiliation{School of Computational Sciences, Korea Institute for Advanced Study, Seoul 02455, Korea}
\author{Deok-Sun \surname{Lee}}
\email{deoksunlee@kias.re.kr}
\affiliation{School of Computational Sciences, Korea Institute for Advanced Study, Seoul 02455, Korea}
\affiliation{Center for AI and Natural Sciences, Korea Institute for Advanced Study, Seoul 02455, Korea}

\begin{abstract}
We study the structure of the international trade hypergraph consisting of triangular hyperedges representing the exporter-importer-product relationship. Measuring the mean hyperdegree of the adjacent vertices, we first find its behaviors different from those in the pairwise networks and explain the origin by tracing the relation between  the hyperdegree and the pairwise degree.  To interpret the observed hyperdegree correlation properties in the context of trade strategies, we decompose the correlation into two components by identifying one with the background correlation remnant even in the exponential random hypergraphs preserving the given empirical hyperdegree sequence.  The other component characterizes the net correlation and reveals the bias of  the exporters of low hyperdegree towards the importers of high hyperdegree and the products of low hyperdegree, which information is not readily accessible in the pairwise networks. Our study demonstrates the power of the hypergraph approach in the study of real-world complex systems and offers a theoretical framework. 
\end{abstract}
\date{\today}
\maketitle 

\section{Introduction}
Given the high complexity and enormous scale of various real-world complex systems, network science~\cite{Albert2002,Newman2010} takes their simplest representation - network - capturing the global connectivity patterns of elements and explores the emergent dynamical behaviors, advancing for the past decades  our knowledge in social networks~\cite{Watts1998,Wasserman1994}, biological and ecological networks~\cite{Jeong2000,Dunne2002}, airline route networks~\cite{Guimera2005,Wuellner2010,Zanin2013}, technological information networks~\cite{Faloutsos1999,Albert1999} and so on. 
The connectivity of elements is often mapped with pairwise edges, for which it is assumed that  higher-order interactions involving more than two elements are infrequent or mediated indirectly by a succession of pairwise interactions.  Yet group interactions do occur in many systems including a group of people communicating simultaneously, co-authors publishing a paper, multiple  proteins forming complexes, and the living species under group interactions. Their impact can be so crucial as to be relevant to the emergent dynamics like phase transitions and scaling in e.g., synchronization and spreading~\cite{Battiston2020,Battiston2021}. The simultaneous connection  of more than two elements is represented by a $d$-clique of $d>2$ elements and  the study of higher-order networks such as simplicial complexes~\cite{Courtney2016,Petri2018,Costa2016,Courtney2016,Zuev2015} from algebraic topology~\cite{Hatcher2002} or hypergraphs~\cite{Ghoshal2009,Zlatic2009, Jhun2019,Stasi2014} has attracted much attention recently.

Compared with the attention paid to and the following rapid development of the theory of higher-order networks, its application to real-world systems has been relatively poor, impeding further development of the field.  It is  partly due to the lack of large-scale empirical data containing the full information of higher-order interactions except for a few cases like the co-authorship networks~\cite{Lee2021}.  In this paper, we study the global organization of international trade by applying the hypergraph approach. Empirical trade data-sets~\cite{Gleditsch2002,Feenstra2005,WTO} provide the annual values of the distinct product categories exported by countries to destination countries. An individual trade is specified by the exporter country, the importer country, and the product, and their collection can be represented by a hypergraph consisting of triangular hyperedges connecting an exporter, an importer, and a product. So far most studies have considered the pairwise networks in the projected space including the country-country networks~\cite{Serrano2003,Garlaschelli2004,Oh2017}, the product-product networks~\cite{Hidalgo2007,Barbier2017}, and the country-product networks~\cite{Bustos2012,Lei2015,Choi2019,*Choi2021,Saracco2015}. 

The landscape of international trade may look disordered as can be expected from the heterogeneous distribution of production resources and different economic, social, and cultural circumstances of countries.   Yet  the trade hypergraphs exhibit a robust connectivity pattern informing us of the trade strategies of countries.  To detect a bias in forming triangular (exporter-importer-product) hyperedges, we measure the mean hyperdegree of the adjacent vertices -~importer or product~-  of  an exporting country as a function of the hyperdegree of the latter as has been studied for pairwise networks~\cite{Satorras2001,Newman2002,Garlaschelli2004, Saracco2015}. We find that as the hyperdegree of an exporter increases, the adjacent  importer's hyperdegree decreases but the adjacent product's one stays constant. This result is different from the corresponding correlation properties reported for the pairwise trade networks, and we identify the origin in tracing the information loss in projecting hypergraphs onto pairwise networks. Next, to translate correctly the observed hyperdegree correlations into the trade strategies of countries, we compare them with those of an ensemble of random hypergraphs. Specifically, we  introduce the exponential random hypergraphs (ERH), which are maximally random for a given hyperdegree sequence, and show analytically and numerically that the strong heterogeneity of vertex hyperdegrees induces even in the ERH a negative correlation between the hyperdegrees of adjacent vertices, which we call the background correlation. Differences from the background correlations are the net correlations of the real trade hypergraphs, which allow us to identify true biases  in selecting partner countries and products. Modifying the ERH model, we construct a model for correlated hypergraphs, which reproduces the empirical net correlations and helps better understand their nature. 

The paper is organized as follows. In Sec.~\ref{sec:empirical}, we construct the trade hypergraphs and measure the mean hyperdegrees of the nearest neighbors to characterize the hyperdegree correlation properties. In Sec.~\ref{sec:ERH}, we introduce the ERH model and present its properties, which provide the background correlation and allow us to identify the net correlations of the trade hypergraphs. In Sec.~\ref{sec:CH}, a model for correlated hypergraphs is proposed. We summarize and discuss our findings in Sec.~\ref{sec:conclusion}. 

\section{Empirical trade hypergraphs}
\label{sec:empirical}

\subsection{Construction of a trade hypergraph}

\begin{figure}[t]
\centering
\includegraphics[width=\columnwidth]{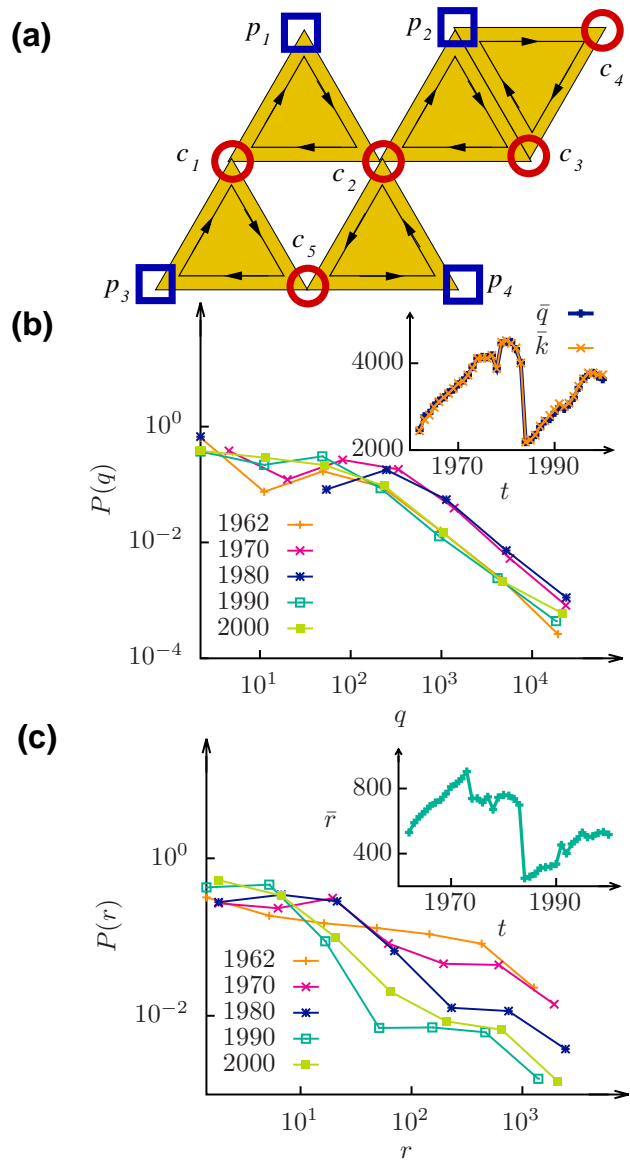}
\caption{
An example of trade hypergraph and the empirical hyperdegree distributions. (a) A trade hypergraph of 5 countries (circles), 4 products (squares), and 5 hyperedges (triangles). Arrows represent the relationship of connected vertices.  For instance, a country $c_1$ exports a product $p_1$ to a country $c_2$. 
(b) The distribution of the export hyperdegree of a country in selected years. Inset: The mean export hyperdegree $\overline{q}$ and the mean import hyperdegree $\overline{k}$ versus time (year). 
(c) The distribution of the trade hyperdegree of a product. Inset: The mean trade hyperdegree $\overline{r}$ versus time. 
}
\label{fig:stat}
\end{figure}

The NBER-UN data-sets~\cite{Gleditsch2002,Feenstra2005}  contain the information of all the traded products, classified by the standard international trade classification (SITC), along with their exporting and destination countries aggregated over each year $t$ in the period $1962\leq t\leq 2000$.  Each annual trade can be best represented by a triangular hyperedge connecting an exporter, an importer, and a traded product, the whole collection of which yields the trade hypergraph as exemplified in Fig.~\ref{fig:stat}(a). Two types of vertices, country and product, are present in the trade hypergraphs, denoted by $c$ and $p$ respectively.  Each triangular hyperedge is directed as characterized by clockwise or anti-clockwise cyclic arrows. The arrow from a country $c$ to a product $p$ means that $c$ exports $p$ while  the arrow from a product $p$ to a country $c$ means that $c$ imports $p$. The arrow from a country $c_1$ to another country $c_2$ means that $c_1$ imports something  from  $c_2$.

The adjacency tensor $a_{cc'p}(t)$ for each 3-tuple $(c,c',p)$ represents whether a country $c$ exports a product $p$ to another country $c'$ or not in year $t$, i.e., 
\begin{equation}
    a_{cc'p}(t)=
    \begin{cases}
    1 & {\rm if} \ c \ {\rm exports} \ p \ {\rm to} \ c' \ {\rm in \ year} \ t,\\
    0 & {\rm otherwise}.
    \end{cases}
\label{eq:adj_hyper}
\end{equation}
While the trade values are also available~\cite{Squartini2011,*Squartini2011_2,Bhattacharya2008,Almog2015,Barbier2017,Oh2017,Choi2019,*Choi2021} and one can construct the weighted trade hypergraphs, it is beyond the scope of the present study and we focus on the unweighted version.
We consider two vertices as being adjacent to or the nearest neighbor of each other if they are connected by an hyperedge, like e.g., $c_1$ and $c_2$ located at two vertices of a triangular hyperedge in Fig.~\ref{fig:stat}(a). Note that for $a_{cc'p}$, the first index indicates the exporter country, the second  the importer country, and the final one  the traded product.

\subsection{Hyperdegrees and their broad distributions}
\label{sec:hyperdeg}

In the trade hypergraphs, the number of hyperedges attached to a vertex represents the total number of distinct trades involving it.  Distinguishing the role and type of each vertex, we define  three kinds of hyperdegrees as 
\begin{equation}
q_c \equiv \sum_{c'p}\, a_{cc'p}, \ \ k_{c'} \equiv \sum_{cp} a_{cc'p}, \ \ 
r_p \equiv \sum_{cc'}a_{cc'p}. 
\label{eq:hyperdegree}
\end{equation}
The export hyperdegree $q_c$ of a country $c$ represents the number of the distinct pairs of a destination (importer country) $c'$ and a traded product $p$ appearing in the export portfolio $\{(c',p)|a_{cc'p}>0\}$ of $c$, quantifying its diversification. The import hyperdegree $k_{c'}$ of a country $c'$ characterizes the diversification of the import portfolio $\{(c,p)|a_{cc'p}>0\}$ of $c'$. Finally the trade hyperdegree $r_p$ of a product $p$ can be a measure of the popularity of $p$ in international trade.

An hyperedge $h = (c,c',p)$ consists of three edges, $e=(c,p), \, e'=(c,c')$, and $e''=(c,'p)$, and we will characterize the correlations of the two vertices connected by one of the three edges of an hyperedge in terms of the hyperdegrees of the vertices as given in Eq.~(\ref{eq:hyperdegree}). If we are given one such edge, e.g., $(c,p)$, then we can connect a vertex, say $c'$, to the edge to complete the formation of an hyperedge $(c,c',p)$. The principles underlying the latter procedure also deserve investigation though it is beyond the scope of the present work.  As introduced in Ref.~\cite{Zlatic2009}, one can define the hyperdegrees of edges,  e.g., $S_{(c,p)} \equiv \sum_{c'}\, a_{cc'p}$ and $U_{(c,c')} \equiv \sum_p a_{cc'p}$, which can be helpful for characterizing the correlations of an edge and a vertex connected by the same hyperedge.  In Appendix~\ref{sec:pairwise} and \ref{sec:bipartite}, we briefly introduce the properties of the hyperdegrees of edges and the vertex-edge networks, which are bipartite networks of vertices and edges representing the connection of edges and vertices.

The hyperdegrees  of vertices are found to be broadly distributed throughout the whole period [Figs.~\ref{fig:stat}(b) and \ref{fig:stat}(c)]. Broad degree distributions have been identified also in the pairwise trade networks~\cite{Serrano2003,Garlaschelli2004}. Such heterogeneous hyperdegrees can be attributed to different natural and social environments of countries and different distributions of available production resources of products, and thus one can consider the hyperdegrees of countries and products as their characteristic capacity. The mean hyperdegrees, $\overline{q} = L/\Nex$, $\overline{k} = L/\Nim$, and $\overline{r} = L/\Npr$, with $L\equiv \sum_{cc'p}a_{cc'p}$ the total number of hyperedges and $\Nex, \Nim, \Npr$ the numbers of exporting countries, importing countries, and traded products, vary with time, which reflects the growth or recession of the global economy despite the possibility of the  incompleteness of the data-sets~\cite{Gleditsch2002}. 

To characterize quantitatively the way of connecting vertices by hyperedges, we investigate the correlation of the hyperdegrees of the nearest-neighbor vertices, as has been extensively studied for various complex networks~\cite{Satorras2001,Newman2002} including trade networks~\cite{Garlaschelli2004,Saracco2015}.

\subsection{Mean hyperdegree of the nearest-neighbor vertices: Hyperdegree correlation}

Heterogeneous hyperdegrees of countries and products represent their different capacity to be connected to {\it any} vertices and constrain strongly the organization of the trade hypergraphs.  To detect the organization principles taken beyond  the given hyperdegree sequence, one should investigate which countries and which product are connected and what the associated probability is.  In this light we measure the  mean hyperdegree of the vertices adjacent to a vertex and study its dependence of the hyperdegree of the latter to detect their correlations. To be specific, let us consider for an exporting country $c$
\begin{align}
\overline{\kcnn} &= \frac{\sum_{c'p} k_{c'} \, a_{cc'p}}{\sum_{c'p} a_{cc'p}} = { \sum_{c'p} k_{c'} \, a_{cc'p} \over q_c}, \ {\rm and} \ \nonumber\\
\overline{\rcnn} &= \frac{\sum_{c'p} r_{p} \, a_{cc'p}}{\sum_{c'p} a_{cc'p}} = {\sum_{c'p} r_p \, a_{cc'p} \over q_c}, 
\label{eq:knncrnnc}
\end{align}
where `nn' indicates `nearest neighbor' or `adjacent'.  They are the mean import hyperdegree of the countries importing from and the mean trade hyperdegree of the products exported by the country $c$, respectively. They can vary with the export hyperdegree $q_c$, which can be probed by $\overline{\knn}  (q) = {\sum_c  q_c \,\overline{\kcnn} \, \delta (q_c - q) \over \sum_c q_c \, \delta(q_c - q)} = \frac{\sum_{cc'p} k_{c'} \, a_{cc'p} \, \delta(q_c-q)}{\sum_{cc'p} a_{cc'p} \, \delta(q_c-q)}$ and  $\overline{\rnn }(q) = {\sum_c q_c \, \overline{\rcnn} \, \delta (q_c - q) \over \sum_c q_c \, \delta(q_c - q)} = \frac{\sum_{cc'p} r_{p} \, a_{cc'p}\, \delta(q_c-q)}{\sum_{cc'p} a_{cc'p} \, \delta(q_c-q)}$.

One can expect that $\overline{\kcnn}$ and $\overline{\rcnn}$ will be independent of $c$ if vertices are connected randomly. If it is the case, $\overline{\kcnn}$ and $\overline{\rcnn}$ are reduced to the edge-based mean hyperdegrees 
\begin{align}
\overline{\knn} &= {\sum_{cc'p} k_{c'} a_{cc'p} \over \sum_{cc'p} a_{cc'p}} = {\overline{k^2}\over \overline{k}}, \ {\rm and} \ \nonumber\\
     \overline{\rnn} &= {\sum_{cc'p} r_p a_{cc'p} \over \sum_{cc'p} a_{cc'p}}= {\overline{r^2}\over \overline{r}}, 
\label{eq:knnrnn}
\end{align}
respectively.  In general, these local and global mean hyperdegrees of the nearest neighbors in Eqs.~(\ref{eq:knncrnnc}) and (\ref{eq:knnrnn}) respectively  are related to each other by $\overline{\knn} = \sum_c q_c  \overline{\kcnn}/L$ and $\overline{\rnn} = \sum_c q_c \overline{\rcnn}/L$.  On the other hand, if $\overline{\kcnn}$ and $\overline{\rcnn}$ vary with $c$ or equivalently $\overline{\knn}(q)$ and $\overline{\rnn}(q)$ vary with $q$, they will deviate from the global means $\overline{\knn}$ and $\overline{\rnn}$, pointing out the bias that a country has in selecting its partner importer countries and export products.

\begin{figure}[t]
\centering
\includegraphics[width=\columnwidth]{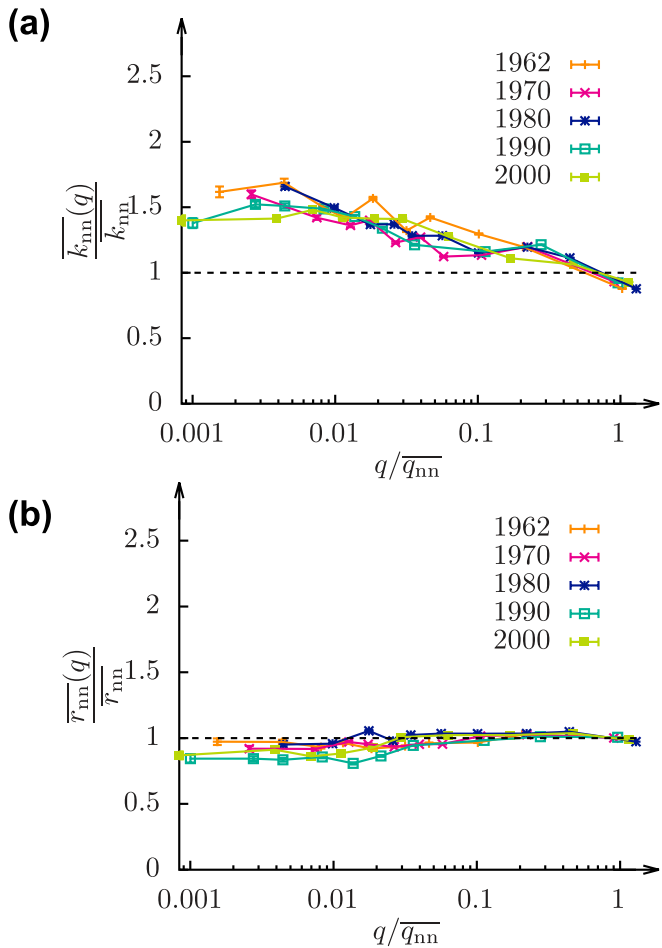}
\caption{
The mean hyperdegrees of the nearest-neighbor vertices in the trade hypergraphs.
(a) The mean import  hyperdegree $\overline{\knn}(q)$ of the countries importing from a country of export  hyperdegree $q$ scaled by the edge-based mean import hyperdegree $\overline{\knn}$ versus the scaled export  hyperdegree $q/\overline{\qnn}$ in selected years.   
(b) The mean trade  hyperdegree $\overline{\rnn}(q)$ of the products exported by a country of export  hyperdegree $q$ scaled by the edge-based mean trade hyperdegree $\overline{\rnn}$ versus  $q/\overline{\qnn}$.
}
\label{fig:moment_empirical}
\end{figure}

It turns out that the ratio $\overline{\knn}(q)/\overline{\knn}$ decreases with increasing $q$, from about $1.6$ to $0.9$ [Fig.~\ref{fig:moment_empirical}(a)]. It implies that  a country of a low export hyperdegree and equivalently a weakly diversified export portfolio tends to select an importing country of a high import hyperdegree. In contrast, $\overline{\rnn}(q)/\overline{\rnn}$ is almost constant [Fig.~\ref{fig:moment_empirical}(b)], suggesting that a country selects randomly a product to export, independent of its popularity. These empirical behaviors are one of the main results of the present study, but their interpretations need caution as will be shown soon. 

In Appendix~\ref{sec:pairwise}, we project the trade hypergraphs onto the subspace in which either of importers or  products are neglected, and obtain the pairwise trade networks~\cite{Serrano2003,Garlaschelli2004,Hidalgo2007,Bustos2012,Lei2015,Barbier2017,Saracco2015,Oh2017,Choi2019,*Choi2021}. Both the mean degree of the importers adjacent to an exporting country in the exporter-importer (E-I) networks and that of the products adjacent to an exporter in the exporter-product (E-P) networks decrease with the degree of the exporter suggesting negative correlations~\cite{Choi2019,*Choi2021,Saracco2015}. The hypergraph results in Fig.~\ref{fig:moment_empirical} are different from those for the pairwise networks, particularly regarding the behaviors of $\overline{\rnn}(q)$. We show in Appendix~\ref{sec:pairwise} that the relation between the hyperdegree and the pairwise degree of a product varies with which exporters prefer it; the products exported preferentially by the countries of low export hyperdegree (or degree) are exported by many distinct countries but their destinations are not so diverse as for the products exported by the countries of high export hyperdegrees. This causes the correlation properties to appear different between the hypergraphs and the pairwise networks.

In addition to the comparison with the degree correlations in the pairwise networks, the significance of the observed hyperdegree correlations should be assessed to infer the underlying trade strategies of countries.  It has been shown that without any explicit correlation or bias, the heterogeneity of degrees can generate a correlation in the degrees of adjacent vertices due to the exclusion of multiple edges~\cite{Park2003,*Park2004}, which has been applied also to the trade networks~\cite{Saracco2015}. Given the broad hyperdegree distributions shown in Figs.~\ref{fig:stat}(b) and \ref{fig:stat}(c), the same can happen also in the trade hypergraphs, and we should investigate how much of the measured values of $\overline{\knn}(q)$ and $\overline{\rnn}(q)$ are contributed to by such {\it background} correlation originating from the degree heterogeneity and how large the remaining {\it net} correlation is, the latter of which will show us the true bias of countries in composing their export portfolios. To this end,  we introduce the maximally random hypergraphs preserving the empirical hyperdegree sequences and analyze their hyperdegree correlation properties to compare with the empirical ones in the next sections.

 \section{Exponential random hypergraphs: Background and net correlation}
 \label{sec:ERH}

Following Ref.~\cite{Stasi2014,Park2003,*Park2004, Saracco2015}, we  consider the ensemble of hypergraphs with the probability measure of each hypergraph $G = (a_{cc'p})$ given by $P(G) = Z^{-1} e^{H(G)}$ and the graph Hamiltonian $H(G)$ given in the form
\begin{equation}
    H(G) = \sum_{cc'p} (\thetaex_c + \thetaim_{c'}+\thetapr_p) a_{cc'p}.
         \label{eq:HG}
\end{equation}
The Gibbs entropy $S = -\langle \ln P\rangle = -\sum_G P(G) \ln P(G)$ is maximized with this probability measure under the constraints that the expectation values of hyperdegrees should be equal to prescribed ones $q_c, k_{c'} , r_p$ and the expected number of links be equal to $L$ as
\begin{align}
\sum_{c'p} \langle a_{cc'p}\rangle &= q_c, \ \ 
\sum_{cp} \langle a_{cc'p}\rangle = k_{c'}, \nonumber\\
\sum_{cc'} \langle a_{cc'p}\rangle &= r_p, \ \ 
\sum_{cc'p} \langle a_{cc'p}\rangle = L
\label{eq:average}
\end{align}
 with the notation $\langle \cdots\rangle$ used to denote the ensemble average $\langle F \rangle = \sum_G P(G) F(G)$.
The Lagrange multipliers $\thetaex_c$, $\thetaim_{c'}$, and $\thetapr_p$ in Eq.~(\ref{eq:HG}) play the role of inverse temperatures and are determined by Eq.~(\ref{eq:average}).  The hypergraphs with the probability measure given by   Eq.~(\ref{eq:HG}) can be called the exponential random hypergraph (ERH) after Ref.~\cite{Park2003,*Park2004}.  As the probability measure is factorized as $P(G) = Z^{-1}\prod_{cc'p} e^{(\thetaex_c + \thetaim_{c'}+\thetapr_p) a_{cc'p}}$ and the adjacency tensor element $a_{cc'p}$ is either $0$ or $1$,  one can evaluate the partition function $Z \equiv \sum_G e^{H(G)}$ as 
$Z = \prod_{cc'p} \sum_{a_{cc'p}=0}^1 e^{(\thetaex_c + \thetaim_{c'}+\thetapr_p)a_{cc'p}}  = \prod_{cc'p} \left(1 + e^{\thetaex_c + \thetaim_{c'}+\thetapr_p}\right)$, and 
 the expected adjacency tensor  $\langle a_{cc'p}\rangle = {\partial \log Z \over \partial (\thetaex_c + \thetaim_{c'}+\thetapr_p) } $ is given by  
\begin{align}
    \langle a_{cc'p}\rangle &= \aERH_{cc'p} \equiv {e^{\thetaex_c + \thetaim_{c'}+\thetapr_p} \over 1+e^{\thetaex_c + \thetaim_{c'}+\thetapr_p}} \nonumber\\
    &= {\Lambda \, \lambdaex_c \lambdaim_{c'} \lambdapr_p \over 1 + \Lambda \, \lambdaex_c \lambdaim_{c'} \lambdapr_p},
    \label{eq:aERH}
\end{align}
where we introduce the normalized export, import, and trade {\it fitness} $\lambdaex_c \equiv {e^{\thetaex_c} \over \sum_{c'} e^{\thetaex_{c'}}}, \lambdaim_c \equiv {e^{\thetaim_c} \over \sum_{c'} e^{\thetaim_{c'}}}, \lambdapr_p \equiv {e^{\thetapr_p} \over \sum_{p'} e^{\thetapr_{p'}}}$ of countries and products and the connection fitness $\Lambda \equiv \sum_{cc'p} e^{\thetaex_c + \thetaim_{c'} + \thetapr_p}$. The values of $\lambda$'s and $\Lambda$ can be considered as the fitness in gaining hyperedges of each vertex and the whole hypergraph.  Solving Eq.~(\ref{eq:average}) with Eq.~(\ref{eq:aERH}), one can obtain all the fitness values. Notice that the probability measure is represented also as 
\begin{equation}
P(G) = \prod_{cc'p} \left( {\aERH_{cc'p} }\right)^{a_{cc'p}} \left(1 - \aERH_{cc'p}\right)^{1-a_{cc'p}}
\label{eq:PGERH}
\end{equation}
with the parameters $\aERH_{cc'p}$ given in Eq.~(\ref{eq:aERH}) being the expectation value of $a_{cc'p}$ or the probability of $c,c'$ and $p$ to be connected. 

Once the fitness values are given, the expected adjacency tensor $\aERH_{cc'p}$ in Eq.~(\ref{eq:aERH}) is fixed for every tuple $(c,c',p)$ of vertices.  Then we can construct a hypergraph by connecting every tuple $(c,c',p)$ with probability $\aERH_{cc'p}$ or leaving the tuple disconnected with probability $1-\aERH_{cc'p}$.  In this way one can create a realization of the ERH for given fitness values and  can obtain as many realizations as possible by repeating this procedure.   

It should be noted that the ERH preserves the empirical hyperdegree sequence, by Eq.~(\ref{eq:average}), and therefore the moments of the hyperdegree distributions are preserved in the ERH, e.g., $\overline{\knn^{\rm ERH}} = \overline{\knn} = {\overline{k^2}\over \overline{k}}$ and $\overline{\rnn^{\rm ERH}} = \overline{\rnn} = {\overline{r^2}\over \overline{r}}$ while   the hyperdegrees of the nearest neighbors may be changed as they depend on which tuples of vertices are connected.

\subsection{Sparse limit}

The specific form of the adjacency tensor as given in Eq.~(\ref{eq:aERH}) is related to the constraint that $\aERH_{cc'p}$ cannot be larger than one even for large $\lambda$'s or $\Lambda$. In the limit in which $\Lambda \lambdaex_c \lambdaim_{c'} \lambdapr_p \ll 1$ for all $c,c',p$,  the adjacency tensor is approximated as $\aERH_{cc'p}\simeq  \Lambda \lambdaex_c \lambdaim_{c'} \lambdapr_p$, and one finds the fitness values proportional to the prescribed vertex degrees or the total number of links as
\begin{equation}
    \lambdaex_c \simeq {q_c \over L}, \ \    \lambdaim_{c'} \simeq  {k_{c'} \over L}, 
    \lambdapr_p \simeq {r_p \over L}, \ \ 
    \Lambda \simeq L
    \label{eq:lambdaS}
\end{equation}
from Eq.~(\ref{eq:average}).
These results help understand the relation between degree and fitness in the ERH intuitively. Using Eq.~(\ref{eq:lambdaS}), we find  
\begin{equation}
\aERH_{cc'p}\simeq {q_c k_{c'} r_p \over {L}^2}.
\label{eq:aS}
\end{equation}
A necessary condition for the above limit to be valid is that 
\begin{equation}
    L {\overline{q} \over L} { \overline{k} \over L} {\overline{r} \over L}   = {L \over \Nex \Nim \Npr} = {L\over L_{\rm max}}\ll 1
\end{equation}
meaning that the empirical number of hyperedges $L$ should be much smaller than its maximum possible value, $L_{\rm max}=\Nex \Nim \Npr$, and thus the  hypergraph is sparse.   Empirically the trade hypergraphs  have a low edge density $\ell\equiv L/L_{\rm max}\simeq 0.02$, leading us to expect Eq.~(\ref{eq:aS}) to be valid.  

Inserting Eq.~(\ref{eq:aS}) into Eq.~(\ref{eq:knncrnnc}), one can obtain the mean hyperdegrees of the nearest-neighbor vertices $\overline{\kcnn^{\rm ERH}}$ and $\overline{\rcnn^{\rm ERH}}$ for the ERH and find them to be equal to the edge-based mean degrees $\overline{\knn}$ and $\overline{\rnn}$, respectively,  independent of $c$ in the sparse limit. Recalling that the real trade hypergraphs  have $\overline{\knn}(q)$ decreasing with $q$ while $\overline{\rnn}(q)$ constant approximately,  one can suspect that  the ERH does not lie in the true sparse limit where Eqs.~(\ref{eq:lambdaS}) and (\ref{eq:aS}) are valid, and/or the real trade hypergraphs are different from the ERH but connected in a non-random way. We will soon show that both are the case.  
In the next subsection we first show that the strong heterogeneity of hyperdegrees drives the ERH out of the sparse regime although the edge density is low. 

\subsection{Deviation from the sparse limit due to hyperdegree heterogeneity}

\begin{figure}
\centering
\includegraphics[width=\columnwidth]{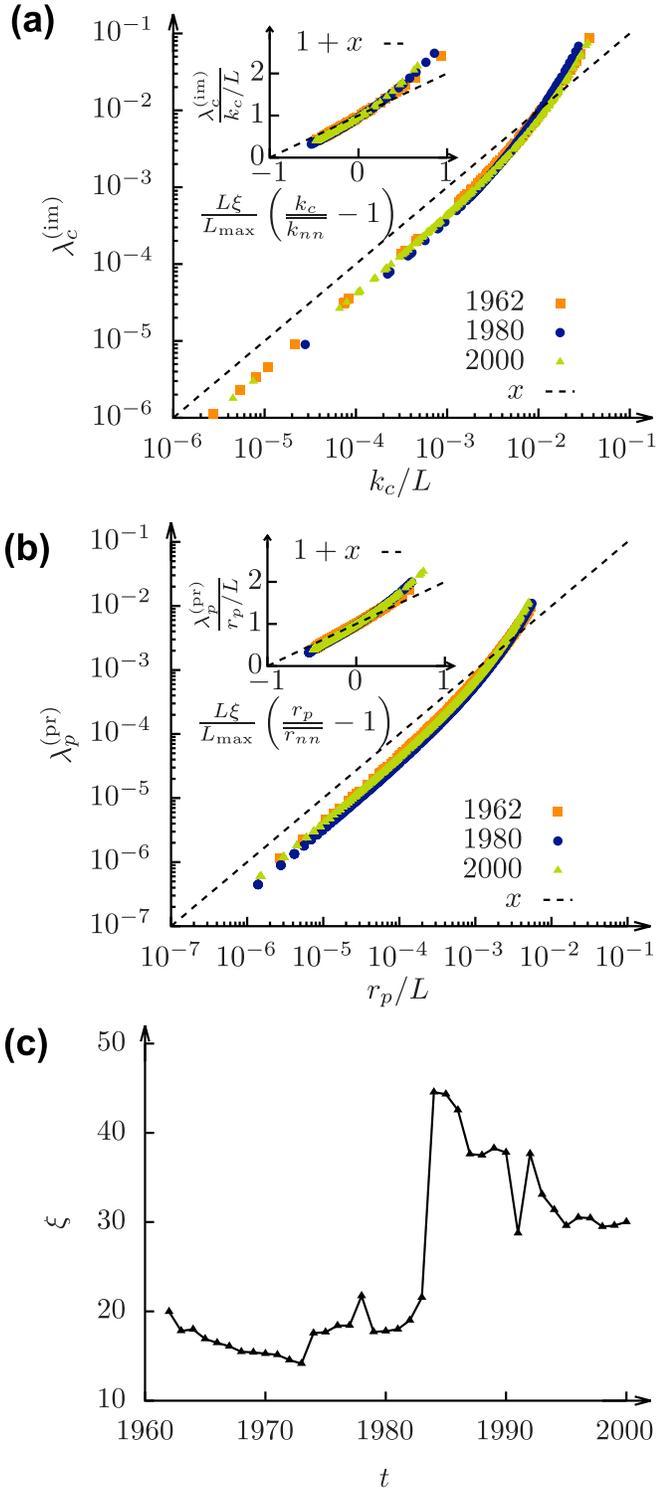}
\caption{
Fitness of countries and products and hyperdegree heterogeneity in the ERH for international trade. 
(a) The import fitness versus the scaled import hyperdegree of countries. $L$ is the total number of hyperedges. The dashed line represents the sparse-limit prediction. Inset: Deviation from the sparse limit behavior versus the scaled import degree from Eq.~(\ref{eq:lambda1}).  
(b) The trade fitness versus the scaled trade hyperdegree of products. Inset:  Deviation from the sparse limit behavior. 
(c) Hyperdegree heterogeneity versus time (year).
}
\label{fig:ERH}
\end{figure}

The fitness values of vertices for the ERH obtained by solving Eq.~(\ref{eq:average}) and Eq.~(\ref{eq:aERH}) with the empirical hyperdegree sequences are presented  in Fig.~\ref{fig:ERH}. The behaviors of the fitness of countries and products as functions of their hyperdegrees look similar but different from  the sparse-limit prediction [Fig.~\ref{fig:ERH} (a) and (b)]; the fitness values  are smaller for vertices of low hyperdegrees and larger for those of high hyperdegrees than the sparse-limit predictions $k_c/L$ and $r_p/L$ as given in Eq.~(\ref{eq:lambdaS}). Such deviations  induce hyperdegree correlations even in the ERH.

To understand the origin of the deviations, we solve Eq.~(\ref{eq:average}) when the edge density $\ell= L/L_{\rm max}$ is low by expanding the fitness variables in terms of $\ell$ as e.g., $\lambdaim_c = \lambdaim_{c,0} + \lambdaim_{c,1}\ell + \lambdaim_{c,2}\ell^2 + \cdots$ and $\Lambda = \Lambda_1 \ell + \Lambda_2 \ell^2 + \Lambda_3 \ell^3+ \cdots$. The calculations are straightforward and we find up to the first order in $\ell$ that  
\begin{align}
\Lambda &= L\left[1 + \ell\, \xi +O\left(\ell^2 \right)\right],\nonumber\\
\lambdaex_{c} &= {q_c \over L} \left\{1 + \ell \, \xi \left({q_c \over \overline{\qnn}}  - 1\right) + O\left(\ell^2 \right)\right\},\nonumber\\
\lambdaim_{c'} &= {k_{c'} \over L} \left\{1 + \ell \, \xi \left({k_{c'} \over \overline{\knn}}  - 1\right) + O\left(\ell^2 \right)\right\},\nonumber\\
\lambdapr_{p} &= {r_p \over L} \left\{1 + \ell \, \xi \left({r_p \over \overline{\rnn}}  - 1\right) + O\left(\ell^2 \right)\right\},
\label{eq:lambda1}
\end{align}
where $\overline{\qnn} = {\overline {q^2 }\over \overline{q}}$ is the edge-based mean export hyperdegree defined similarly to Eq.~(\ref{eq:knnrnn}). 
 The deviation of the fitness from the zeroth-order results in Eq.~(\ref{eq:lambdaS}) can be driven not only by a  dense edge density but also by a large value of the hyperdegree heterogeneity $\xi$ defined as
\begin{equation}
\xi \equiv {\overline{\qnn} \over \overline{q}} {\overline{\knn} \over \overline{k}} {\overline{\rnn} \over \overline{r}} = {\overline{q^2} \over \overline{q}^2}{\overline{k^2} \over \overline{k}^2}{\overline{r^2} \over \overline{r}^2}.
\label{eq:xi}
\end{equation}
It is the product of the ratios of the second moment to the square of the first moment for the hyperdegree distributions. Each ratio, e.g., ${\overline{q^2}\over \overline{q}^2}$, would be slightly larger than one, as it is given by $1 + {1\over \overline{q}}$ and $\overline{q}$ is quite large [Fig.~\ref{fig:stat}(a)], if $P(q)$ were the Poisson distribution. The empirical hyperdegree distributions are broader than the Poisson distributions as  seen in Fig.~\ref{fig:stat}, leading  to $\overline{q^2}/\overline{q}^2$ fluctuating about 5, $\overline{k^2}/\overline{k}^2$ about 2, and $\overline{r^2}/\overline{r}^2$ about 2. Therefore the hyperdegree heterogeneity $\xi$ characterizes the broadness of all the hyperdegree distributions, quantifying the heterogeneity of hyperdegrees. It ranges roughly between 10 and 50 empirically  though fluctuating with time [Fig.~\ref{fig:ERH}(c)], contributing to the deviation of the ERH from the sparse limit. Inserting Eq.~(\ref{eq:lambda1}) into Eq.~(\ref{eq:aERH}), we see that
\begin{align}
\aERH_{cc'p} &= {q_c k_{c'} r_p \over L^2} \left\{1 - \ell \, \xi \left( {q_c \over \overline{\qnn}} {k_{c'} \over \overline{\knn}} {r_p \over \overline{\rnn}} \right.\right.\nonumber\\
&-\left.\left. {q_c \over \overline{\qnn}} - {k_{c'} \over \overline{\knn}} - {r_p \over \overline{\rnn}} +2 \right) + O(\ell^2)\right\}.
\label{eq:a1}
\end{align}

One can consider Eqs.~(\ref{eq:aS}) and (\ref{eq:lambdaS}) as the zeroth-order results in the limit $\ell\, \xi \to 0$, in which the exclusion of multiple edges play little role.  As $\ell\, \xi $ increases,  the edges that would connect multiply hub vertices  are used to connect non-hub vertices, which results in the fitness of hub (non-hub) vertices larger (smaller) than the sparse-limit prediction as given in the first-order correction term in Eq.~(\ref{eq:lambda1}). The criterion of whether the fitness is larger or smaller than the sparse-limit prediction is given by the edge-based mean trade hyperdegree $\overline{\knn}$ or $\overline{\rnn}$ as supported in Figs.~\ref{fig:ERH}(a) and \ref{fig:ERH}(b).

The exponential random graph approach can be applied generally. In Appendix~\ref{sec:bipartite}, we present the graph Hamiltonian and the fitness values for the exponential random vertex-edge graphs that will be defined therein.

\subsection{Background and net correlations}

\begin{figure}[t]
\centering
\includegraphics[width=\columnwidth]{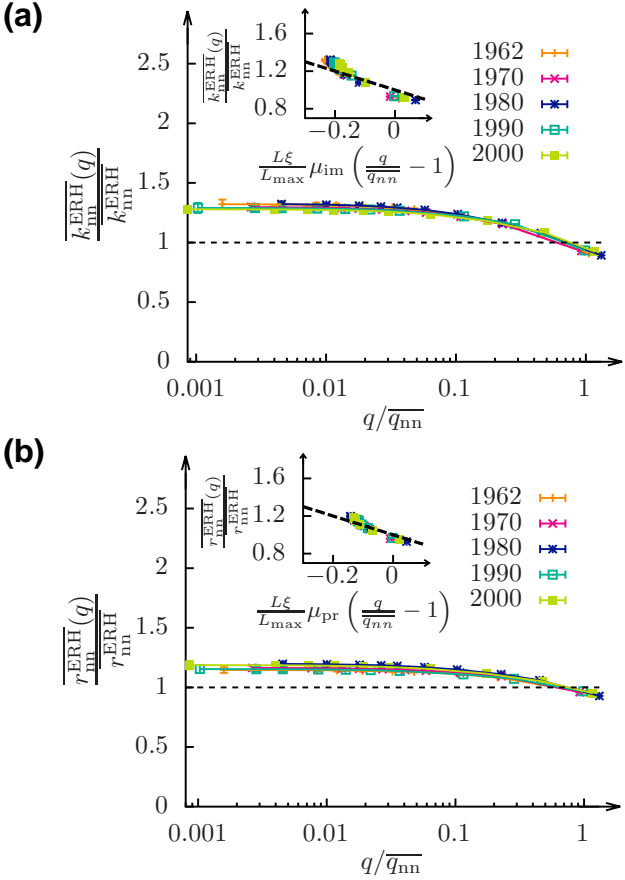}
\caption{
Background correlations. 
(a) The scaled mean import  hyperdegree $\overline{\knn^{\rm ERH}}(q)/\overline{\knn^{\rm ERH}}$ of the countries importing from a country of export  hyperdegree $q$ in the ERH for selected years. Note that $\overline{\knn^{\rm ERH}} = \overline{\knn}$. Inset: The same plot as a function of the scaled export hyperdegree as predicted by  Eq.~(\ref{eq:knncrnncanalytic}) with the dashed line for $y=1-x$. Here $\mu_{\rm im} = {\overline{k_{\rm com}}\over \overline{\knn}}-1 = {\overline{k^3} \, \overline{k} \over \overline{k^2}^2}-1$. 
 (b) The scaled mean trade hyperdegree $\overline{\rnn^{\rm ERH}}(q)/\overline{\rnn^{\rm ERH}}$ of the products exported by a country of export hyperdegree $q$ in the ERH. Note that $\overline{\rnn^{\rm ERH}} = \overline{\rnn}$. Inset: The same plot as a function of the scaled export hyperdegree with $\mu_{\rm pr} = {\overline{r_{\rm com}}\over \overline{\rnn}}-1 = {\overline{r^3} \, \overline{r} \over \overline{r^2}^2}-1$ and the dashed line for $y=1-x$. 
 }
\label{fig:moment_ERH}
\end{figure}

As the ERH of international trade is not within the sparse limit [Fig.~\ref{fig:ERH}], the  mean hyperdegrees of the nearest-neighbor vertices in the ERH  vary with $q$. Inserting the numerically-obtained fitness $\lambda$'s  into Eq.~(\ref{eq:aERH}) to obtain $\aERH_{cc'p}$ for all $c,c',p$ and using the obtained $\aERH_{cc'p}$ in place of $a_{cc'p}$ in Eq.~(\ref{eq:knncrnnc}), one can obtain $\overline{\kcnn^{\rm ERH}}$ and $\overline{\rcnn^{\rm ERH}}$, which are shown in Figs.~\ref{fig:moment_ERH}(a) and ~\ref{fig:moment_ERH}(b).  

Both $\overline{\knn^{\rm ERH}}(q)$ and $\overline{\rnn^{\rm ERH}}(q)$ remain larger than the sparse-limit predictions $\overline{\knn}$ and $\overline{\rnn}$, respectively, for small $q$ and decrease with $q$ in the whole range of $q$,  commonly meaning negative correlations of the hyperdegrees of adjacent vertices.   Such $q$-dependent behaviors can be understood analytically by inserting Eq.~(\ref{eq:a1}) into Eq.~(\ref{eq:knncrnnc}) to obtain the approximate expressions up to the first order in the edge density as 
\begin{align}
{\overline{\kcnn}\over \overline{\knn}}&=  1 - \ell \, \xi \, \left({\overline{k_{\rm com}}  \over \overline{\knn}}-1\right) \left({q_c \over \overline{\qnn}}-1 \right) + O(\ell^2),\nonumber\\
{\overline{\rcnn} \over \overline{\rnn}} &=  1 - \ell \, \xi \, \left({\overline{r_{\rm com}}  \over \overline{\rnn}}-1\right) \left({q_c \over \overline{\qnn}}-1 \right) + O(\ell^2),
\label{eq:knncrnncanalytic}
\end{align}
where $\overline{k_{\rm com}} \equiv {\sum_{c_1, c_2, c, p_1,p_2} a_{c_1cp_1} k_c a_{c_2 cp_2} \over \sum_{c_1, c_2, c, p_1,p_2} a_{c_1cp_1}  a_{c_2 cp_2}} = {\overline{k^3}\over \overline{k^2}}$ and $\overline{r_{\rm com}} \equiv {\overline{r^3}\over \overline{r^2}}$ are the mean import and trade hyperdegree of the common neighbors of two exporting countries, respectively.  As $\overline{k^3}\overline{k} - \overline{k^2}^2 = \sum_{c,c'} k_c k_{c'} (k_c - k_{c'})^2/(2\Nim^2)>0$, both the factors $\mu_{\rm im} \equiv  {\overline{k_{\rm com}}  \over \overline{\knn}}-1$ and $\mu_{\rm pr} \equiv  {\overline{r_{\rm com}}  \over \overline{\rnn}}-1$ are positive and thus one can expect that $\overline{\knn^{\rm ERH}}(q)$ and $\overline{\rnn^{\rm ERH}}(q)$ to decrease linearly with $q$, which is confirmed in the insets of Fig.~\ref{fig:moment_ERH}(a) and ~\ref{fig:moment_ERH}(b). As discussed in the previous subsection,  such negative hyperdegree correlations in the ERH do not imply any bias in connecting vertices but are generated by assigning edges just to avoid connecting multiply the same tuples, most likely those exhibiting ${q_c k_{c'} r_p \over L^2}\gg 1$ which may occur frequently when the edge density $\ell$ and the hyperdegree heterogeneity $\xi$ are large. Therefore the correlation identified in the ERH can be considered as the background correlation in that  it is generated even without any explicit bias in connecting vertices.  

\begin{figure}[t]
\centering
\includegraphics[width=\columnwidth]{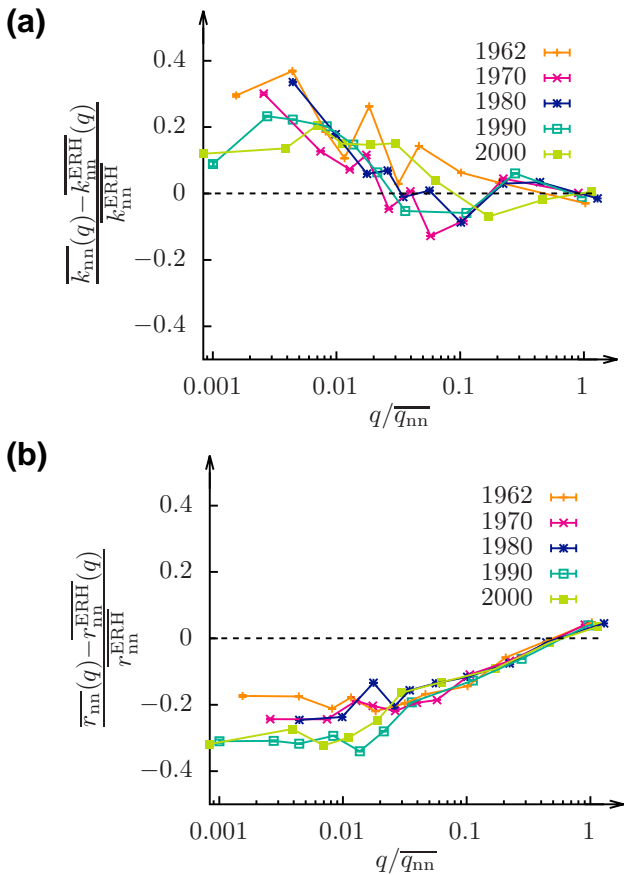}
\caption{
Net correlations.
 (a)  The net component $\overline{\knn^{\rm NET}}(q)  = \overline{\knn}(q) - \overline{\knn^{\rm ERH}}(q)$ of the mean import hyperdegree of the nearest-neighbor importers scaled by the edge-based mean import hyperdegree $\overline{\knn^{\rm ERH}}$ of the ERH is shown for selected years.  
 (b) The net component $\overline{\rnn^{\rm NET}}(q)  = \overline{\rnn}(q) - \overline{\rnn^{\rm ERH}}(q)$ of the mean trade hyperdegree of the nearest-neighbor products scaled by $\overline{\rnn^{\rm ERH}}$.
}
\label{fig:moment_NET}
\end{figure}

The empirical hyperdegree correlations display extra deviations from the background ones. The mean hyperdegree ${\overline{\knn}(q)}$ of the nearest-neighbor importers decreases more steeply with $q$ in the real trade hypergraphs [Fig.~\ref{fig:moment_empirical}(a)]  than ${\overline{\knn^{\rm ERH}}(q)}$ of the ERH [Fig.~\ref{fig:moment_ERH}(a)]. Most remarkably, the mean trade hyperdegree ${\overline{\rnn}(q)}$ of the nearest-neighbor products  varies little with $q$ empirically [Fig.~\ref{fig:moment_empirical}(b)] while ${\overline{\rnn^{\rm ERH}}(q)}$ decreases with $q$ in the ERH [Fig.~\ref{fig:moment_ERH}(b)].  Such differences lead us to decompose the mean hyperdegree of the nearest neighbors into the background and the net component as
\begin{align}
\overline{\kcnn} &=\overline{\kcnn^{\rm ERH}} +  \overline{\kcnn^{  \rm NET}}, \ {\rm and} \  \nonumber \\
\overline{\rcnn} &=\overline{\rcnn^{\rm ERH}}+  \overline{\rcnn^{ \rm NET}}.
\end{align}
The $q$-dependent net components $\overline{\knn^{\rm NET}}(q)$ and $\overline{\rnn^{\rm NET}}(q)$ are shown in Figs.~\ref{fig:moment_NET}(a) and \ref{fig:moment_NET}(b), which are another main results of the present study.

The net components reveal  different principles that exporting countries take in selecting parter importers and  products to export. The positivity of $\overline{\knn^{  \rm NET}}(q)$ for small $q$ suggests that the countries with low export hyperdegree are more likely to select the countries with high import hyperdegrees than expected  in the ERH, i.e., based on their hyperdegrees only. It decreases  with increasing $q$ to be zero or even negative, implying that such bias in the parter selection is weakened or reversed for the countries having diversified export portfolios. In contrast, $\overline{\rnn^{\rm NET}}(q)$ is negative for small $q$, which suggests that the countries of low export hyperdegree are likely to export less popular products than expected in the ERH.  Moreover, it increases with $q$; such negative bias is weakened with increasing $q$ towards zero or a positive value.  As discussed in Appendix~\ref{sec:pairwise}, the latter bias is related to the fact that the products exported by the countries of low export hyperdegrees tend to have few destinations and thus have low hyperdegrees.
On the other hand the exporters with diversified portfolios export products that have diverse destinations and thus have higher hyperdegrees than expected in the ERH. 

Our interest turns to  designing a model for the correlated hypergraphs reproducing the empirical net correlations. If such a model is available, its difference from the ERH will enable a deeper understanding of  the structural organization of the real-world trade hypergraphs. In the next section, we construct the correlated hypergraph model by modifying the ERH model. 

\section{A model for correlated hypergraphs}
\label{sec:CH}

 The model hypergraphs reproducing the observed net correlations beyond the ERH can help better understand the nature of the obtained degree correlations and locate the real-world trade hypergraphs in the appropriate parameter space. As they should have both the background correlations and the net correlations, we consider a modification of the ERH.
 
Let us consider the ensemble of graphs for which the probability measure $P(G)$ is given as in Eq.~(\ref{eq:PGERH}) with the parameters $\aERH_{cc'p}$'s replaced by  
\begin{equation}
\aCH_{cc'p} 
= \min \left\{1, \aERH_{cc'p} \Normc  {(\lambdaex_c + \lambdaim_{c'})^\alpha \over (\lambdaex_c + \lambdapr_{p})^\beta }\right\}.
\label{eq:aCH}
\end{equation}
and $\Normc$ being the normalization constant satisfying $q_c  = \sum_{c'p}\aCH_{cc'p}$ for each country $c$. Note that  $\aCH_{cc'p}$ is the expectation value of the adjacency tensor element, i.e.,  $\langle a_{cc'p}\rangle = \aCH_{cc'p}$ in the model and that  the fitness values $\Lambda$ and $\lambda$'s obtained in the ERH are used. The correction term  $\acorr_{cc'p} \equiv  \Normc {(\lambdaex_c + \lambdaim_{c'})^\alpha \over (\lambdaex_c + \lambdapr_{p})^\beta }$ is introduced to reproduce the net correlations at the cost of inducing deviations of the hyperdegrees of individual vertices between data and model. The exponents $\alpha$ and $\beta$ are estimated such that they yield the correlation properties as much close as possible to the empirical ones.  We will call the hypergraphs constructed by this model the correlated hypergraphs (CH). Connecting each tuple of $(c,c',p)$ with probability $\aCH_{cc'p}$ in Eq.~(\ref{eq:aCH}), we can generate a realization of the CH model. 

 The specific form of the correction term $\acorr_{cc'p}$  is motivated by the $q$-dependence of  $\overline{\knn^{ \rm NET}}(q)$ and $\overline{\rnn^{  \rm NET}}(q)$ shown in Figs.~\ref{fig:moment_NET}(a) and \ref{fig:moment_NET}(b). Let us first examine the dependence of $\acorr_{cc'p}$ on $\lambdaim_{c'}$  for $\alpha>0$. The correction term will increase with $\lambdaim_{c'}$ when $\lambdaex_c \ll \lambdaim_{c'}$ while it will be almost constant, independent of $\lambdaim_{c'}$, when $\lambdaex_{c}\gg \lambdaim_{c'}$. Recalling that $\lambdaex_c$ and $\lambdaim_{c'}$  grow with increasing $q_c$ and $k_{c'}$, respectively [Fig.~\ref{fig:ERH} (a)], one can expect that a country $c$ with low $q_c$ will be connected more preferentially to the importing countries $c'$ of high $k_{c'}$ than expected in the ERH, due to $\acorr_{cc'p}$. On the other hand a country $c$ with high $q_c$ will be connected to other countries in the same manner as expected in the ERH, which is in agreement with the empirical net correlations shown in Fig.~\ref{fig:moment_NET}(a).  Similarly,  for $\beta>0$,  $\acorr_{cc'p}$ will decrease with $\lambdapr_{p}$ only when $\lambdaex_c \ll \lambdapr_{p}$, leading us to expect  a country $c$ of low $q_c$ to be connected preferentially to the products $p$ of low $r_p$ compared with the ERH prediction as observed empirically in  Fig.~\ref{fig:moment_NET} (b).  

\begin{figure}[t]
\centering
\includegraphics[width=\columnwidth]{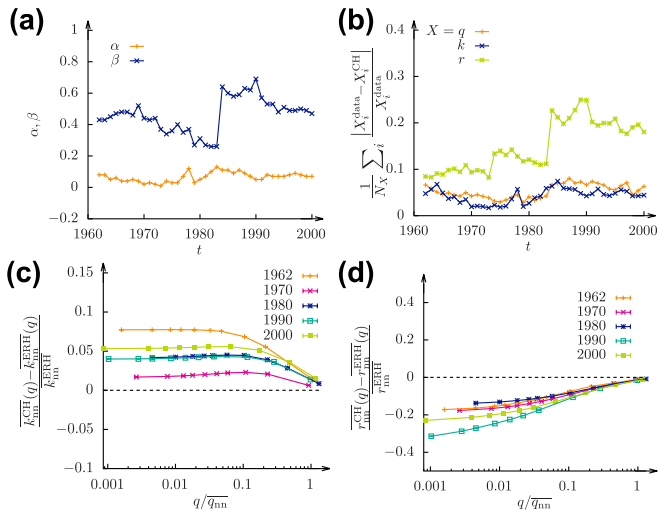}
\caption{CH model and net correlations.
(a) The estimated exponents $\alpha$ and $\beta$ of the CH model as functions of time.
(b) The relative error  of the hyperdegrees of vertices.
(c)   The net component $\overline{\knn^{\rm CH}}(q) - \overline{\knn^{\rm ERH}}(q)$ of the mean import hyperdegree of the nearest-neighbor importers in the CH scaled by $\overline{\knn^{\rm ERH}}$ are shown for selected years.  
 (d) The net component $\overline{\rnn^{\rm CH}}(q)  - \overline{\rnn^{\rm ERH}}(q)$ of the mean trade hyperdegree of the nearest-neighbor products in the CH  scaled by $\overline{\rnn^{\rm ERH}}$.
}
\label{fig:CH}
\end{figure}

To estimate the exponents $\alpha$ and $\beta$, we  substitute $\aCH_{cc'p}$ in Eq.~(\ref{eq:knncrnnc}) to evaluate the mean hyperdegrees of the adjacent vertices $\overline{\kcnn^{\rm CH}}$  and $\overline{\rcnn^{\rm CH}}$ for every vertex $c$ and minimize the differences from the empirical values quantified by 
\begin{equation}
\Enn (\alpha, \beta) = \sum_{c, q_c>0} \left| \overline{\kcnn^{\rm CH}} - \overline{\kcnn}\right| \sum_{c, q_c>0} \left| \overline{\rcnn^{\rm CH}} - \overline{\rcnn}\right| .
\label{eq:error}
\end{equation}
The estimated values of $\alpha$ and $\beta$ are both positive, as expected, and the latter is larger; $\alpha$ and $\beta$ fluctuate with time around $\alpha\simeq 0.07$ and $\beta\simeq 0.5$, respectively [Fig.~\ref{fig:CH}(a)].

Owing to the introduction of the correction term $\acorr_{cc'p}$ to the adjacency tensor of the ERH,  the hyperdegrees are necessarily deviating from the empirical values  [Fig.~\ref{fig:CH}(b)]. Nevertheless the export and import hyperdegrees remain close to the empirical values, due to the introduction of $\Normc$ and the small value of $\alpha$, respectively while the trade hyperdegrees show significant, 10 to 20\%, deviations. We find that the net component of the mean hyperdegree of the nearest-neighbor importers $\overline{\knn^{\rm CH}}(q) - \overline{\knn^{\rm ERH}}(q)$ remains positive and decreases with $q$ though the variation and fluctuation with time is weaker than in the real trade hypergraphs [Fig.~\ref{fig:CH}(c)]. The net component for  the nearest-neighbor products  $\overline{\rnn^{\rm CH}}(q) -\overline{\rnn^{\rm ERH}}(q)$ remains negative and increases with $q$ in  quite good agreement with the empirical behaviors [Fig.~\ref{fig:CH}(d)]. 

It is remarkable that such simple models as proposed in Eq.~(\ref{eq:aCH}) reproduce the empirical characteristics of the net correlations, suggesting the possibility of constructing simple models for other complex hypergraphs. The CH model can be viewed as a first-order approximation towards reproducing the empirical net correlations; we used the fitness values of the ERH without revising them. Therefore it is desirable to tune $\lambda$'s and $\Lambda$ as well as $\alpha$ and $\beta$ towards satisfying Eq.~(\ref{eq:average}) as well as minimizing Eq.~(\ref{eq:error}) in a future research. Our model can be a guide in devising the models capturing the structural characteristics of various real-world hypergraphs and facilitating systematic studies.   

\section{Summary and discussion}
\label{sec:conclusion}

The trade hypergraphs that we have investigated consist of the hyperedges representing the ternary relationship of a product, an exporter, and  an importer, and thus capture the full information of international trade compared with the pairwise networks representing the binary relationship.  Defining three kinds of hyperdegrees,
we have explored the organization principles at the hypergraph level by analyzing the behaviors of the mean hyperdegrees of the nearest neighbors that characterize the correlations of  the hyperdegrees of the adjacent vertices.  We have found different correlation properties  from those known in the pairwise networks and shown that the origin lies in the different ratio of the hyperdegree to the pairwise degree of a product depending on which countries export it preferentially. Taking the correlation remnant in the ERH as the background correlation,  we have identified the net correlations that contain the information of  the true bias that a country has in designing its export portfolio. 

Given many studies on the pairwise networks, our study invokes the importance of hypergraph approach in understanding the organization of real-world complex systems. Lots of structural characteristics  like nestedness  were studied in the pairwise trade networks, and thus it is desirable to extend the studies to the trade hypergraphs.  Also we can take the higher-level descriptions of international trade than presented here for its deeper understanding e.g., by constructing weighted hypergraphs with the hyperedge weights given by the trade values. Also tracing the lost information in projecting the higher-order description to the lower ones, like hypergraphs to weighted or binary pairwise networks, can provide an opportunity to investigate the best description balanced between the cost of data collection and the richness of the information that will be obtained. 

The ERH that we have introduced and analyzed in details can be used for extracting the true non-random features for given heterogeneous hyperdegree sequences of the empirical hypergraphs.  We have also proposed the CH model  that reproduce reasonably the net correlations. Although the model is restricted to a specific data-set, we believe that our methodology to realize the empirical degree correlation via a minimal correction of connecting probability and determine the parameters by minimizing the difference from the empirical data can be applied widely to various types of data-sets.

\begin{acknowledgements}
This work was supported by the National Research Foundation of Korea (NRF) grants funded by the Korean Government (Grant No. 2019R1A2C1003486 (DSL)) and KIAS Individual Grants (Grants No. CG079901 (DSL) and No. CG074101 (SDY)) at Korea Institute for Advanced Study.  The authors thank Kwang-Il Goh for helpful discussions and acknowledge the Center for Advanced Computation in KIAS for providing computing resources. 
\end{acknowledgements}

\appendix 

\section{Hypergraph versus pairwise networks}
\label{sec:pairwise}

Contracting  the adjacency tensor $a_{cc'p}$ in Eq.~(\ref{eq:adj_hyper}) over either of products or importers,  one can obtain the weighted adjacency matrices 
\begin{equation}
\wEI_{cc'} \equiv \sum_p a_{cc'p}, \ {\rm and} \ 
\wEP_{cp} \equiv \sum_{c'} a_{cc'p}
\label{eq:w}
\end{equation}
and the binary adjacency matrices 
\begin{equation}
\aEI_{cc'} = \theta(\wEI_{cc'}), \ {\rm and} \ 
\aEP_{cp} = \theta(\wEP_{cp})
\label{eq:apair}
\end{equation}
to construct the exporter-importer (E-I) and the exporter-product (E-P) networks, respectively~\cite{Choi2019,*Choi2021}. Here $\theta(x)$ is $1$ for $x\geq 0$ and $0$ otherwise, a different notation from the inverse temperatures in Eq.~(\ref{eq:HG}).  The element $\wEI_{cc'}$ represents the number of hyperedges involving $c$ and $c'$ as the exporter and importer vertex in the trade hypergraph, which is equal to $U_{(c,c')}$ mentioned in Sec.~\ref{sec:hyperdeg} and investigated in Appendix~\ref{sec:bipartite},  and is considered as the weight of the link connecting $c$ and $c'$ in the EI network. Similarly $\wEP_{cp}$ corresponds to the number of hyperedges involving $c$ and $p$, equal to $S_{(c,p)}$, and is the link weight in the E-P network.  The vertex degrees in both pairwise networks are defined as 
\begin{align}
&\qEI_c = \sum_{c'} \aEI_{cc'},  \  \kEI_{c'} = \sum_{c} \aEI_{cc'},\nonumber\\
&\qEP_c = \sum_{p} \aEP_{cp}, \  \rEP_p = \sum_{c} \aEP_{cp}.
\label{eq:pairdegree}
\end{align} 
In the context of weighted (pairwise) networks, generalizing the vertex degree, one can consider the vertex strength defined as the sum of the weights of all the links incident on a vertex. From Eq.~(\ref{eq:w}), one can immediately find that for the E-I and the E-P networks  the vertex strength is equal  to one of the hyperdegrees $q_c, k_{c'}$ or $r_p$.

\begin{figure}
\centering
\includegraphics[width=0.9\columnwidth]{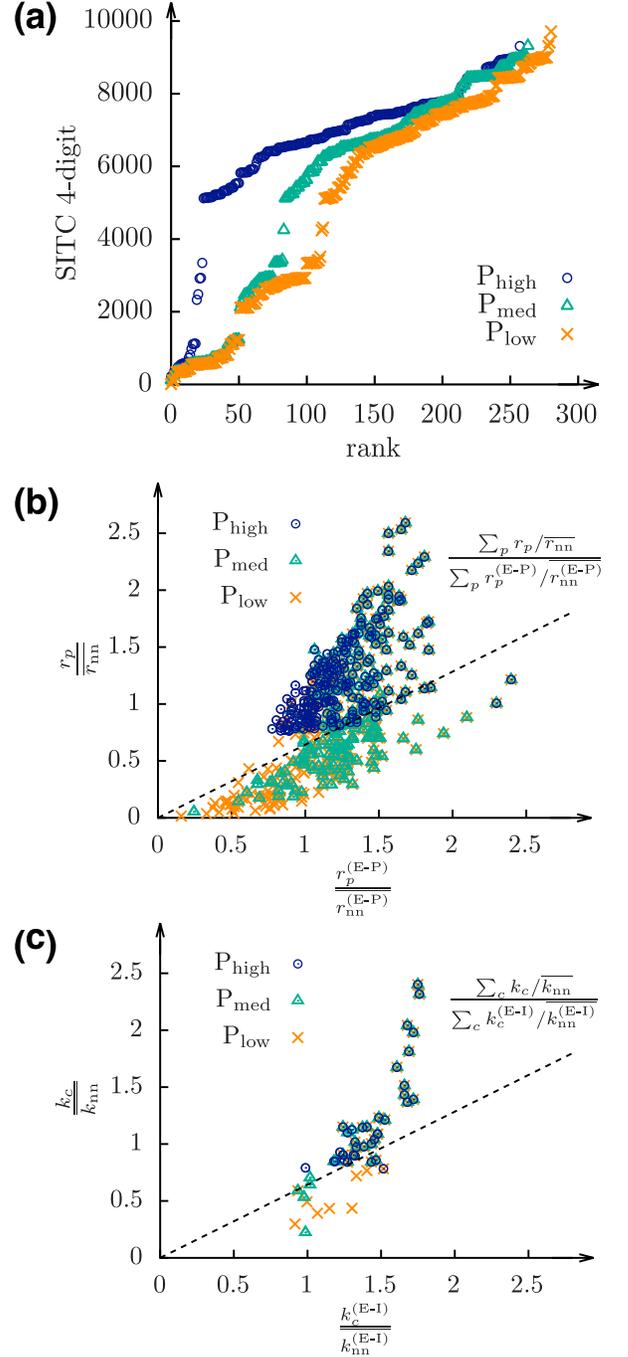}
\caption{Relation between hyperdegrees and degrees. 
(a) SITC codes of the products in three groups ${\rm P}_{\rm high}, {\rm P}_{\rm med}$, and ${\rm P}_{\rm low}$ the top 20\% products preferred by three groups of exporters having high, medium, and low export hyperdegrees  in year 2000. 
(b) The scaled trade hyperdegrees  versus the scaled degrees in the E-P network  for the products preferred by different groups of exporters of different export hyperdegrees.   
(c)  The scaled import hyperdegrees  versus the scaled import degrees in the E-I networks for the importers preferred by different groups of exporters of different export hyperdegrees.  
}
\label{fig:networkprojection}
\end{figure}

The mean degrees of the nearest-neighbor importers and products of an exporting country defined similarly to Eq.~(\ref{eq:knncrnnc}) as 
\begin{align}
\overline{\kcnn^{\rm (E-I)}} &= \frac{\sum_{c} \kEI_{c'} \, \aEI_{cc'}}{\sum_{c'} \aEI_{cc'}} = {1\over \qEI_c} \sum_{c'} \kEI_{c'} \, \aEI_{cc'} , \ {\rm and} \ \nonumber\\
\overline{\rcnn^{\rm (E-P)}} &= \frac{\sum_{p} \rEP_{p} \, \aEP_{cp}}{\sum_{p} \aEP_{cp}} = {1\over \qEP_c} \sum_{p} \rEP_p \, \aEP_{cp}.
\label{eq:knncrnncEIEP}
\end{align}
have been investigated to characterize the connectivity pattern of pairwise trade networks. Previous studies~\cite{Saracco2015,Choi2019,*Choi2021} have consistently shown that both $\overline{\kcnn^{\rm (E-I)}}$ and $\overline{\rcnn^{\rm (E-P)}}$ decrease significantly with $\qEI_c$ and $\qEP_c$, respectively.  It means that the export and import degree of adjacent countries in the E-I network and the degrees of adjacent product and  country  in the E-P network are negatively correlated. Interestingly, such negative correlations are not fully transferred to  the trade hypergraphs; the mean  hyperdegree $\overline{\rcnn}$ of the products exported by an exporting country $c$ appears constant without regard to $q_c$  in the trade hypergraphs [Fig.~\ref{fig:moment_empirical}(b)]. 

Such different behaviors of the mean degree of the nearest-neighbor products between the E-P network and the hypergraph are interesting. If we construct the E-P network and examine the behavior of $\overline{\rcnn^{\rm (E-P)}}$, we will be led to infer that an exporting country of a lower degree, having a smaller number of exporting products, tends to select a product of a higher degree, having a larger number of countries exporting it. On the other hand, when the trade hypergraphs are examined, as in the present study, there is little difference in the trade hyperdegrees of the exported products between the exporting  countries of low and high export hyperdegrees.

The origin of the difference lies in the different definitions and meanings of the vertex degree between the E-P networks and the hypergraphs.  The hyperdegree of a product in the trade hypergraph is related to its degree in the E-P network as
\begin{equation}
r_p = \sum_{cc'} a_{cc'p} = \sum_{c} \wEP_{cp} = \rEP_p \overline{\wEPpnn}
\label{eq:rpreip}
\end{equation}
where $\overline{\wEPpnn} \equiv \sum_{c} \wEP_{cp}/\rEP_p$ is the mean weight of the links incident upon product $p$ in the E-P network, representing the mean number of hyperedges per link. Therefore the hyperdegree $r_p$ is contributed to by the degree $\rEP_p$ and the mean weight of its incident link $\overline{\wEPpnn}$. The latter information  is lost in the binary adjacency matrix  and the vertex degree of the E-P network.
 
The empirical observations are summarized as that the products exported by the countries of low export  hyperdegrees have their network degrees $\rEP_p$ larger than but their hyperdegrees $r_p$ similar to those exported by the countries of high export  hyperdegrees. To understand these, let us divide the exporting countries into three groups $C_{\rm low}, C_{\rm med}$, and $C_{\rm high}$ depending on their export  hyperdegrees $q$ and  identify  the top 20\% ($\sim 200$) products exported by  each group $C$ based on the sum of link weights $\sum_{c\in C}  \wEP_{cp}$. This procedure classifies products into those exported by countries of low, medium, and high export  hyperdegree,  ${\rm P}_{\rm low}, {\rm P}_{\rm med}$, and ${\rm P}_{\rm high}$, respectively,  and allows us to detect how the relation in Eq.~(\ref{eq:rpreip}) varies among them.  The 4-digit SITC codes of these three groups of products are shown in Fig.~\ref{fig:networkprojection}(a). The first digit of the SITC code provides the highest level of classification as follows: 0 for `food and live animals', 1 for `beverages and tobacco', 2 for `crude materials', 3 for `mineral fuels', 4 for `animal and vegetable oils, fats, and waxes', 5 for `chemicals', 6 for `manufactured goods classified by materials', 7 for `machinery and transport equipment', 8 for `miscellaneous manufactured articles', and 9 for `etc'~\cite{Gleditsch2002,Feenstra2005,Choi2019,*Choi2021}. In Fig.~\ref{fig:networkprojection}(a), one can find that the primary products having the first digit  less than $5$ are found more in ${\rm P}_{\rm low}$ while most of the products in ${\rm P}_{\rm high}$ are the manufactured products having the larger first digits~\cite{Choi2019,*Choi2021}.  We find that a product in ${\rm P}_{\rm low}$ tends to have smaller hyperdegree than that in ${\rm P}_{\rm high}$ even when both have the same pairwise degree  in the E-P network [Fig.~\ref{fig:networkprojection}(b)]. In other words, $\overline{\wEPpnn}$ is smaller for the products  in ${\rm P}_{\rm low}$ than those in ${\rm P}_{\rm high}$ and 
such small values of $\overline{\wEPpnn}$ of the former mitigate their large values of $\rEP_p$, resulting in the constancy of $r_p$ among the products preferred by the countries of different export  hyperdegree.  

On the other hand, the relation between the import hyperdegrees $k_c$ and the degree $\kEI_c$ of the importing countries $c$ preferred by each group $C$ of exporters show little variation across different groups of exporters  [Fig.~\ref{fig:networkprojection}(c)], which leads $\overline{\kcnn}$ to decrease with $q_c$ as $\overline{\kcnn^{\rm (E-I)}}$ does with $\qEI_c$.

\section{Vertex-edge networks}
\label{sec:bipartite}

To explore the principles of forming triangular hyperedges in international trade, we have focussed on the pairs of vertices, called adjacent vertices, connected by one of the three edges of each hyperedge. It can be the next step to study how a pair of vertices, located at both ends of an edge, selects the third vertex to complete the formation of an hyperedge. While it  is beyond the scope of the present study and will be investigated elsewhere, here we present basic ideas and some results.

 For the adjacency tensor $a_{cc'p}$ of a trade hypergraph, one can introduce the {\it edge-degrees}~\cite{Zlatic2009}
\begin{equation}
S_{(c,p)} = \sum_{c'} a_{cc'p},  \ {\rm and} \ U_{(c,c')} = \sum_p a_{cc'p}, 
\label{eq:SU}
\end{equation}
and $V_{(c',p)} = \sum_c a_{cc'p}$ as well while we will mainly focus on the two in Eq.~(\ref{eq:SU}). Considering a set of $N_{\rm im}$ importer vertices $I = \{c'| k_{c'}>0\}$ and a set of  $N_{\rm ex,pr}$ (exporter,product) edges $EP=\{e=(c,p)| S_{(c,p)}>0\}$, one can construct a vertex-edge bipartite network with the adjacency matrix $a_{c'e} = a_{cc'p}$ for $c'\in I$ and $e\in EP$, which we will call the I-EP network. Similarly, for a set of $N_{\rm pr}$ product vertices $P = \{p| r_{p}>0\}$ and a set of $N_{\rm ex,im}$ (exporter, importer) edges $EI=\{e'=(c,c')| U_{(c,c')}>0\}$, the P-EI network can be constructed with the adjacency matrix $a_{pe'} = a_{cc'p}$ for $p\in P$ and $e'\in EI$.

\begin{figure}
\centering
\includegraphics[width=\columnwidth]{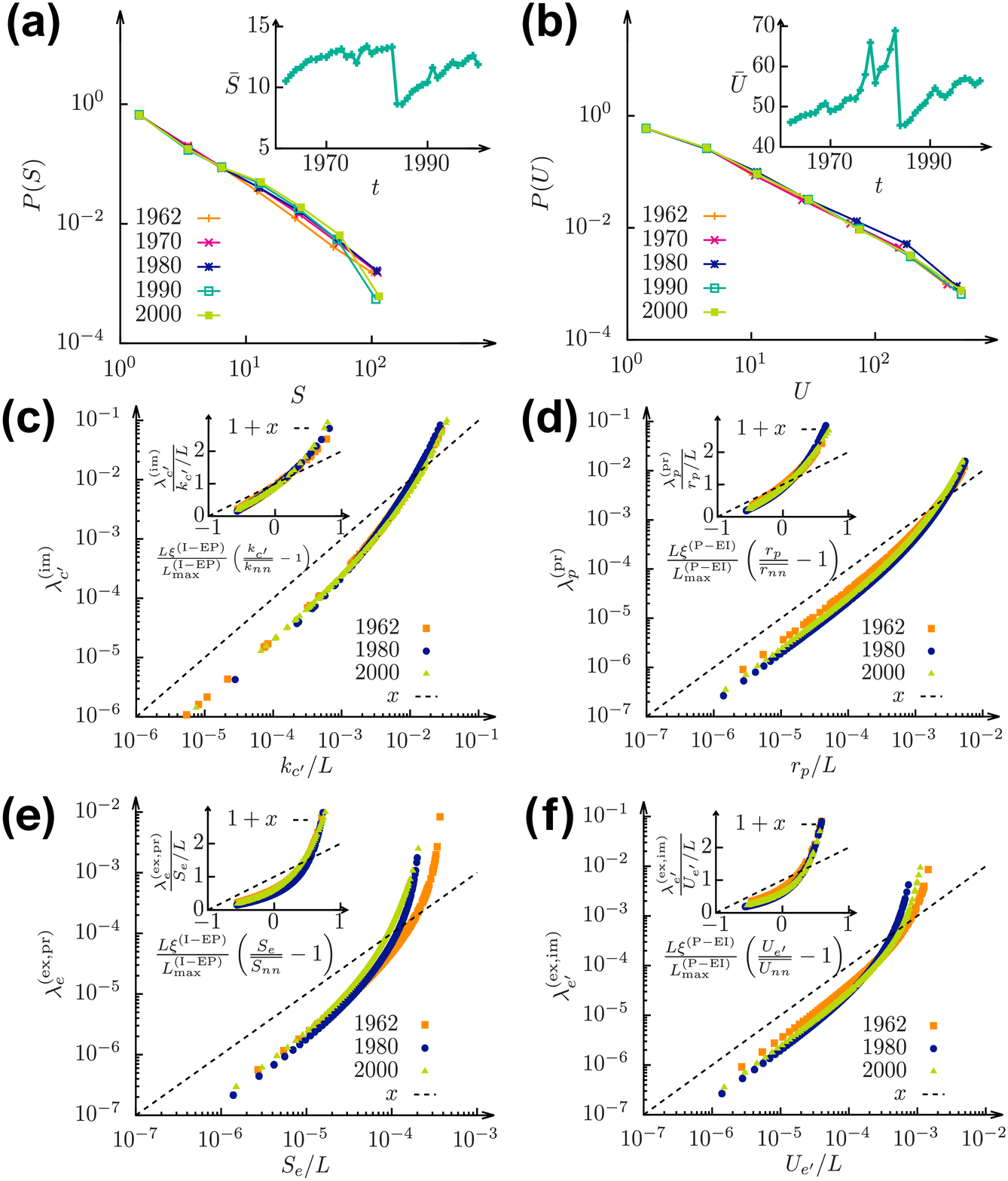}
\caption{(a,b) Empirical edge-degree distributions in selected years. Distributions of (a) the edge-degree $S_{e=(c,p)}$ for the I-EP networks and (b) of $U_{e'=(c,c')}$ for the P-EI networks. Insets: The mean edge-degree versus time. (c-f) Fitness values of the ERVE. Plots of (c) the importer fitness versus the scaled importer degree and (e) the (exporter,product)-edge fitness versus the scaled edge-degree for the I-EP networks. Plots of (d) the product fitness versus the scaled product degree and (f) the (exporter,importer)-edge fitness versus the scaled edge-degree for the P-EI networks.
Insets: Deviations from the sparse limit behavior versus the scaled (edge-)degrees as given in Eq.~(\ref{eq:lambda_approx}).}
\label{fig:bipartite}
\end{figure}

The total number of bipartite edges connecting vertices (in $I$ or $P$) and edges (in $EP$ or $EI$) is equal to the total number of hyperedges in the trade hypergraphs, i.e., $\sum_{c'e} a_{c'e}=\sum_{p e'} a_{p e'} = \sum_{cc'p} a_{cc'p}=L$. Notice also that the degrees of the importer vertices in the I-EP networks and those of the product vertices in the P-EI networks are the same as their corresponding hyperdegrees in the hypergraphs. The edge-degrees are heterogeneous; Both the edge-degree distributions $P(S) = N_{\rm ex,pr}^{-1}\sum_{e\in EP} \delta_{S_e,S}$  and $P(U) = N_{\rm ex,im}^{-1}\sum_{e'\in EI} \delta_{U_{e'},U}$ show power-law tails as shown in Fig.~\ref{fig:bipartite}(a) and ~\ref{fig:bipartite}(b), characterizing the heterogeneous connectivity of the considered vertex-edge bipartite networks.

As  in Sec.~\ref{sec:ERH}, one can construct the exponential random vertex-edge (ERVE) graphs that are maximally random for given degree and edge-degree sequences. Given the constraints for the expectation values of the degrees (of vertices) and the edge-degrees (of edges), the corresponding graph Hamiltonians take the following form
\begin{align}
    H^{\rm (I-EP)}(G=(a_{c'e})) &= \sum_{c', e=(c,p)} (\theta^{\rm (im)}_{c'} + \theta^{\rm (ex,pr)}_{e}) a_{c'e},\ \nonumber\\
    H^{\rm (P-EI)}(G=(a_{pe'})) &= \sum_{p, e'=(c,c')} (\theta^{\rm (im)}_{p} + \theta^{\rm (ex,im)}_{e'}) a_{pe'},
\end{align}
where  $\theta$'s are the Lagrange multipliers. The expected adjacency matrices are represented by
\begin{align}
    \langle a_{c'e}\rangle = a^{\rm ERVE}_{c'e} &= {\Lambda^{\rm (I-EP)} \, \lambda^{\rm (im)}_{c'} \lambda^{\rm (ex,pr)}_{e} \over 1 + \Lambda^{\rm (I-EP)} \, \lambda^{\rm (im)}_{c'} \lambda^{\rm (ex,pr)}_{e} },\ \nonumber\\
     \langle a_{pe'}\rangle = a^{\rm ERVE}_{pe'} &= {\Lambda^{\rm (P-EI)} \, \lambda^{\rm (pr)}_{p} \lambda^{\rm (ex,im)}_{e'} \over 1 + \Lambda^{\rm (P-EI)} \, \lambda^{\rm (pr)}_{p} \lambda^{\rm (ex,im)}_{e'} },
\end{align}
for the I-EP and P-EI networks, respectively. For the I-EP networks, the importer fitness, $\lambda^{\rm (im)}_{c'}={e^{\theta^{\rm (im)}_{c'}}\over \sum_{c} e^{\theta^{\rm (im)}_{c}}}$, the (exporter,product)-edge fitness $\lambda^{\rm (ex,pr)}_{e}={e^{\theta^{\rm (ex,pr)}_{e}}\over \sum_{e'} e^{\theta^{\rm (ex,pr)}_{e'}}}$, and the connection fitness $\Lambda^{\rm (I-EP)}= \sum_{c'e} e^{\theta^{\rm (im)}_{c'} +\theta^{\rm (ex,pr)}_{e}}$ are determined by the constraints $k_{c'}=\sum_e a_{c'e}^{\rm ERVE}$, $S_e = \sum_{c'} a_{c'e}^{\rm ERVE}$, and $L = \sum_{c'e} a_{c'e}^{\rm ERVE}$. Similarly, for the P-EI networks, the fitness values $\lambda^{\rm (pr)}_{p}={e^{\theta^{\rm (pr)}_{p}}\over \sum_{p'} e^{\theta^{\rm (pr)}_{p'}}}$, $\lambda^{\rm (ex,im)}_{e'}={e^{\theta^{\rm (ex,im)}_{e'}}\over \sum_{e} e^{\theta^{\rm (ex,im)}_{e}}}$, and $\Lambda^{\rm I-EP} = \sum_{pe'} e^{\theta^{\rm (pr)}_{p} +\theta^{\rm (ex,im)}_{e'}}$ are determined by the constraints $r_{p}=\sum_{e'} a_{pe'}^{\rm ERVE}$, $U_{e'} = \sum_{p} a_{pe'}^{\rm ERVE}$, and $L = \sum_{p e'} a_{pe}^{\rm ERVE}$.  Solving numerically these equations, we obtain the fitness depending on the corresponding degrees  and edge-edges as seen in Fig.~\ref{fig:bipartite}(c), (d), (e), and (f).

While the total number of bipartite edges is equal to the total number of hypergraphs $L$ both in the I-EP and the P-EI networks, their maximal numbers are different as $L_{\rm max}^{\rm (I-EP)} = N_{\rm im}N_{\rm ex,pr}$ and $L_{\rm max}^{\rm (P-EI)} = N_{\rm im}N_{\rm ex,im}$. One can obtain the fitness values expanded in terms of the bipartite edge densities $\ell^{\rm (I-EP)} = L/L_{\rm max}^{\rm (I-EP)}$  and $\ell^{\rm (P-EI)} = L/L_{\rm max}^{\rm (P-EI)}$  as 
\begin{align}
\lambda_{c'}^{\rm (im)} &\simeq {k_{c'} \over L} \left\{1 + \ell^{\rm (I-EP)} \, \xi^{\rm (I-EP)} \left({k_{c'} \over \overline{k_{\rm nn}}}  - 1\right) \right\},\ \nonumber\\
\lambda_{e}^{\rm (ex,pr)} &\simeq  {S_{e} \over L} \left\{1 + \ell^{\rm (I-EP)} \, \xi^{\rm (I-EP)} \left({S_{e} \over \overline{S_{\rm nn}}}  - 1\right) \right\},\ \nonumber\\
\lambda_{p}^{\rm (pr)} &\simeq  {r_{p} \over L} \left\{1 + \ell^{\rm (P-EI)} \, \xi^{\rm (P-EI)} \left({r_{p} \over \overline{r_{\rm nn}}}  - 1\right) \right\},\ \nonumber\\
\lambda_{e'}^{\rm (ex,im)} &\simeq  {U_{e'} \over L} \left\{1 + \ell^{\rm (P-EI)} \, \xi^{\rm (P-EI)} \left({U_{e'} \over \overline{U_{\rm nn}}}  - 1\right) \right\},
\label{eq:lambda_approx}
\end{align}
where $\xi^{\rm (I-EP)}={\overline{\knn} \over \overline{k}}{\overline{\Snn} \over \overline{S}}$ and  $\xi^{\rm (P-EI)}={\overline{\rnn} \over \overline{r}}{\overline{\Unn} \over \overline{U}}$ characterize the heterogeneity of the I-EP and the P-EI networks, respectively, with $\overline{S_{\rm nn}} = {\overline{S^2} \over \overline{S}}$ and $\overline{U_{\rm nn}} = {\overline{U^2} \over \overline{U}}$.  The numerical solutions to the fitness values qualitatively follow these first-order approximations with some deviations [Fig.~\ref{fig:bipartite} (c-f)].

\bibliographystyle{apsrev4-1}

\end{document}